\documentclass[fleqn,usenatbib]{mnras}
\usepackage{newtxtext,newtxmath}
\usepackage[T1]{fontenc}
\DeclareRobustCommand{\VAN}[3]{#2}
\let\VANthebibliography\thebibliography
\def\thebibliography{\DeclareRobustCommand{\VAN}[3]{##3}\VANthebibliography}

\usepackage{graphicx}
\usepackage{amsmath}
\usepackage{amssymb}
\usepackage{xspace}
\usepackage{listings}

\renewcommand{\vec}[1]{\ensuremath{\mathbf{#1}}\xspace}
\newcommand{\ud}{\ensuremath{\mathrm{d}\xspace}}
\newcommand{\cp}[1]{\texttt{#1}}
\newcommand{\pytransit}{\textsc{PyTransit}\xspace}
\newcommand{\ldtk}{\textsc{LDTk}\xspace}
\newcommand{\swift}{\textsc{RoadRunner}\xspace}
\newcommand{\emcee}{\textsc{emcee}\xspace}
\newcommand{\python}{\textsc{Python}\xspace}

\newcommand{\change}[1]{#1}

\usepackage{scalerel}
\usepackage{tikz}
\usetikzlibrary{svg.path}
\definecolor{orcidlogocol}{HTML}{A6CE39}
\tikzset{
  orcidlogo/.pic={
    \fill[orcidlogocol] svg{M256,128c0,70.7-57.3,128-128,128C57.3,256,0,198.7,0,128C0,57.3,57.3,0,128,0C198.7,0,256,57.3,256,128z};
    \fill[white] svg{M86.3,186.2H70.9V79.1h15.4v48.4V186.2z}
                 svg{M108.9,79.1h41.6c39.6,0,57,28.3,57,53.6c0,27.5-21.5,53.6-56.8,53.6h-41.8V79.1z M124.3,172.4h24.5c34.9,0,42.9-26.5,42.9-39.7c0-21.5-13.7-39.7-43.7-39.7h-23.7V172.4z}
                 svg{M88.7,56.8c0,5.5-4.5,10.1-10.1,10.1c-5.6,0-10.1-4.6-10.1-10.1c0-5.6,4.5-10.1,10.1-10.1C84.2,46.7,88.7,51.3,88.7,56.8z};
  }
}
\newcommand\orcidicon[1]{\href{https://orcid.org/#1}{\mbox{\scalerel*{
\begin{tikzpicture}[yscale=-1,transform shape]
\pic{orcidlogo};
\end{tikzpicture}
}{|}}}}

\title{RoadRunner: a fast and flexible exoplanet transit model}

\author[H. Parviainen]{
H. Parviainen,$^{1,2}$\thanks{E-mail:hannu@iac.es}\orcidicon{0000-0001-5519-1391}
\\
$^{1}$Instituto de Astrof\'isica de Canarias (IAC), E-38200 La Laguna, Tenerife, Spain\\
$^{2}$Dept. Astrof\'isica, Universidad de La Laguna (ULL), E-38206 La Laguna, Tenerife, Spain\\
}

\date{Accepted XXX. Received YYY; in original form ZZZ}
\pubyear{2020}

\begin{document}
\label{firstpage}
\pagerange{\pageref{firstpage}--\pageref{lastpage}}
\maketitle

\begin{abstract}
I present \swift, a fast exoplanet transit model that can use any radially symmetric function to model stellar limb 
darkening while still being faster to evaluate than the analytical transit model for quadratic limb darkening by
\citet{Mandel2002}. CPU and GPU implementations of the model are available in the \pytransit transit 
modelling package, and come with platform-independent parallelisation, supersampling, and support for modelling 
complex heterogeneous time series. The code is written in 
\textsc{numba}-accelerated \textsc{Python} (and the GPU model in \textsc{OpenCL}) without \textsc{C} or \textsc{Fortran} 
dependencies, which allows for the limb darkening model to be given as any \python-callable function. Finally,
as an example of the flexibility of the approach, the latest version of \pytransit comes with a numerical limb darkening 
model that uses \ldtk-generated limb darkening profiles directly without approximating them by analytical models.
\end{abstract}


\section{Introduction}

An exoplanet transit model aims to reproduce the photometric signal caused by a planet crossing over the limb-darkened disk of its host star.
Modelling the transit signal would be a simple problem were it nor for stellar limb darkening (LD). A planetary transit over a uniform disk
can be modelled simply by the area of the stellar disk subtracted by the intersection area of the stellar and planetary disks. However,
the stellar surface brightness changes from the centre to the limb of the star (ignoring other effects such as gravity darkening, flares, 
and spots), and the true transit signal equals to the stellar surface brightness integrated over the stellar disk subtracted by the
surface brightness integrated over the area occluded by the planet.

This integration can be carried out analytically for some simple LD models \citep{Mandel2002}, using a series expansion for
increased model flexibility \citep{Gimenez2006}, using model-specific analytical approximations \citep{Maxted2018a}, or using numerical 
integration \citep{Nelson1972,Kreidberg2015,Maxted2016}.
However, the low-order LD models with analytical solutions may fail to reproduce the true stellar surface brightness profile sufficiently well,
and lead to biases in planet characterisation based on transit light curve analysis \citep{Csizmadia2013,Espinoza2015}. The transit model 
based on series expansion for the general LD model by \citet{Gimenez2006} allows in theory for LD models of unlimited complexity, but the 
computation speed becomes quickly prohibitively slow to reach an absolute precision required (or offered) by space-based photometry. The
models based on approximations, such as the one by \citet{Maxted2018a}, can be very powerful in a some volume of parameter space, but may 
become biased outside the parameter space they were designed for. 
In theory, numerical integration allows the use of any LD model with accuracy bound only by the floating point precision.
However, the computation cost of a model based on numerical integration is generally significantly higher than for the analytical models. 

Here I present \swift, a fast transit model that can use any limb darkening model, reach ppm-precision, and still be faster to evaluate
than the analytical transit model for quadratic LD by \citet{Mandel2002}. The model uses one-dimensional numerical integration approach similar
to the \change{\textsc{EBOP} \citep{Popper1981}, \textsc{JKTEBOP} \citep{Southworth2008}, and \textsc{batman} \citep{Kreidberg2015} packages,} 
but separates the integral of the stellar surface brightness occluded by the
planet into a product of the mean occluded surface brightness and the occluded surface area. This separation is beneficial
because the occluded area is cheap to calculate analytically, while the mean stellar surface brightness can be precomputed into a 
relatively low-resolution interpolation table.

The model is implemented in \pytransit v2.1\!\footnote{\url{https://github.com/hpparvi/PyTransit}} \citep{Parviainen2015}.
The implementation is based on \textsc{numba}-accelerated \textsc{Python} code with no \textsc{C} or \textsc{Fortran} dependencies, what makes
the installation of the package trivial and allows for painless parallelisation in computing environments where compiling 
parallelised code could be nontrivial. Also, due to the pure-\textsc{Python} implementation, the limb darkening model can be any 
\textsc{Python} callable that returns the stellar surface brightness as a function of $\mu$.\!\footnote{Where $\mu=\cos\gamma=\sqrt{1-z^2}$, 
$\gamma$ is the foreshortening angle and $z$ is the normalised distance from the centre of the stellar disk.}

I describe the theory behind the model evaluation in all its simplicity in Sect.~\ref{sec:theory}, detail the steps taken to make
the theory into an efficient implementation in Sect.~\ref{sec:implementation}, give examples of the model usage in Sect.~\ref{sec:usage},
discuss the model performance and scalability in Sect.~\ref{sec:performance}, and finish with conclusions and discussion in
Sect.~\ref{sec:conclusions}.

\section{Theory}
\label{sec:theory}

\begin{figure}
	\centering
	\includegraphics[width=\columnwidth]{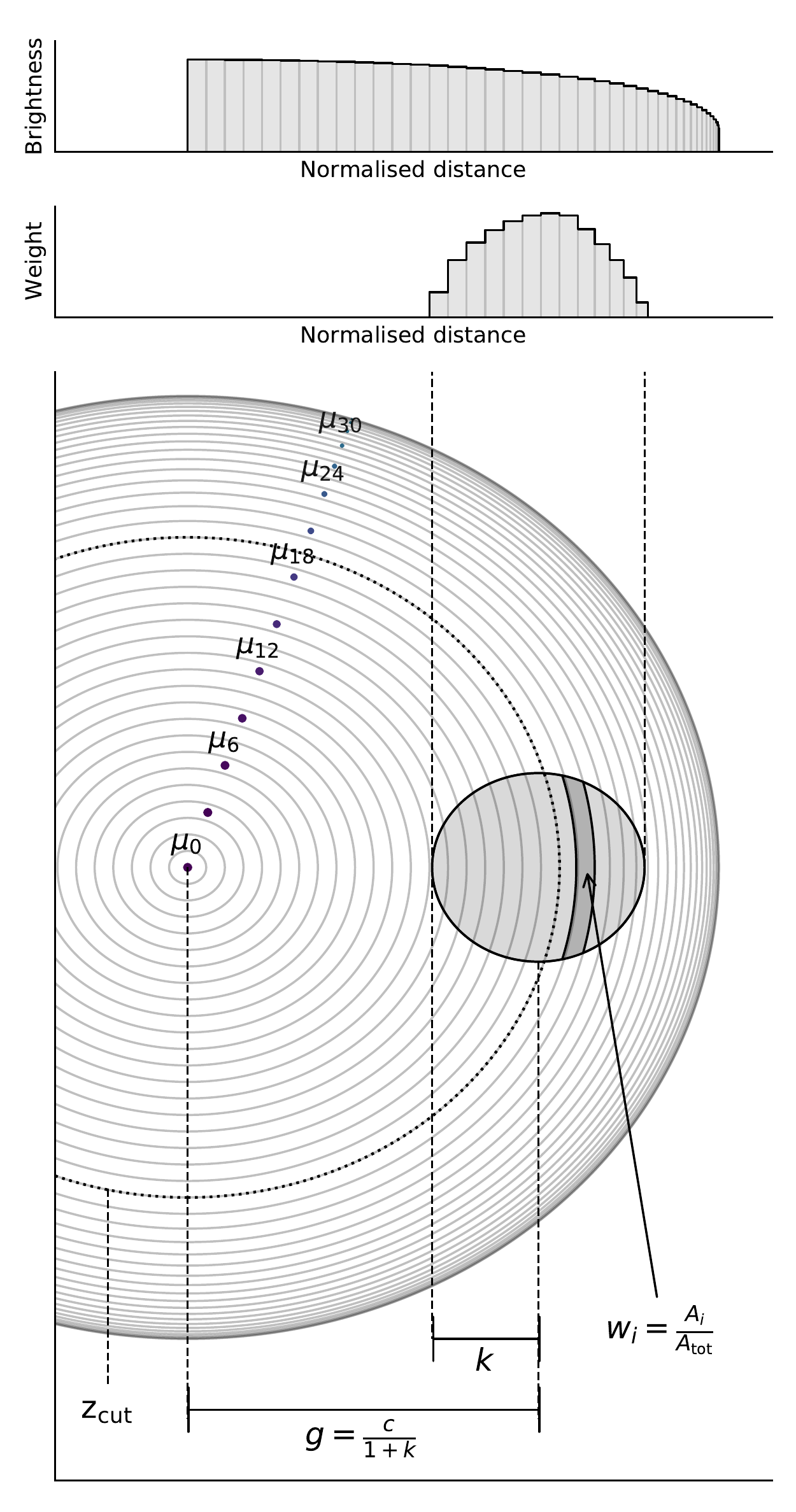}
	\caption{The stellar surface brightness (limb darkening) profile evaluated for an array of distances from the centre of the stellar disk 
	(top), a weight array calculated for a planet with a radius ratio $k$ at a planet-star centre separation $c$ (middle), and a schematic
	showing the stellar disk, planet, the discretisation of the stellar disk, and the calculation of the weight array weights (bottom). The 
	surface brightness and weight arrays are named $\vec{l}$ and $\vec{w}$ in Eq.~\eqref{eq:model}, respectively. 
	}
	\label{fig:ptmodel}
\end{figure}

The transit signal is caused by a planet occluding a part of the limb-darkened stellar surface. 
If the stellar surface brightness can be presented as a radially symmetric function of a normalised distance from the stellar centre, 
$z$, (that is, if we can ignore 
gravitation darkening and other effects breaking the radial symmetry) the transit signal is
\begin{align}
    I &= \int_{A_\star} I(z) \;\ud A - \int_{A_\mathrm{p}} I(z) \;\ud A \\
      &= 2\pi  \int_0^1 z\;I(z)\;\ud z -  \int_{A_\mathrm{p}} I(z) \;\ud A,
\end{align}
where $A_\star$ stands for the area of the star and $A_\mathrm{p}$ stands for the stellar surface occluded by the planet.
The first term can be calculated analytically for most limb darkening models, and efficiently numerically for any model that
might not have an analytical form available. The second term is more complicated since the integration needs to be carried out
over the stellar surface area covered by the planet. However, thanks to the radial symmetry, this integration can also be carried 
out in one dimension, as detailed in \citet{Kreidberg2015}.

The integrals above can also be expressed as products of mean intensities and areas as
\begin{equation}
    I = \hat{I}_\star A_\star - \hat{I}_\mathrm{p} A_\mathrm{p}.  
\end{equation}
Now, by normalising the stellar radius and out-of-transit flux both to unity, we get
\begin{equation}
    I = \frac{\pi\hat{I}_\star - \hat{I}_\mathrm{p} A_\mathrm{p}}{\pi\hat{I}_\star} =  \frac{I_\star - \hat{I}_\mathrm{p} A_\mathrm{p}}{I_\star},
\end{equation}
where $I_\star = \pi \hat{I}_\star$. The area occluded by the planet, $A_\mathrm{p}$, is fast to compute analytically, and the only factor that cannot be 
calculated analytically is the mean surface brightness over the area occluded by the planet.

The mean occluded surface brightness can be estimated given a planet-star radius ratio $k$, the normalised planet-star 
centre distance $c$, and the limb darkening model $l(\mu)$ by first discretising the stellar radius from 0 to 1,\!\footnote{
I discretise the whole stellar radius rather than the radius covered by the planet because it will allow optimisations
discussed later in Sect.~\ref{sec:implementation}}
calculating a weight vector $\vec{w}$ where the weights equal to the areas of the annuli occluded by the planet divided by 
the total area occluded by the planet, and calculating a stellar brightness profile vector $\vec{l}$ by evaluating the
limb darkening model at the bin centres,  as illustrated in Fig.~\ref{fig:ptmodel}. Now, the mean occluded surface brightness 
is given by weighted average that can be computed as a dot product of $\vec{w}$ and $\vec{l}$.

Finally, the transit model for a given $k$, $c$, and limb darkening profile can be evaluated as
\begin{equation}
    I = \frac{I_\star - \left( \vec{w} \cdot \vec{l} \right) \times a(c, k)}{I_\star}, \label{eq:model}
\end{equation}
where $a$ is the circle-circle intersection area.
That is, the transit model evaluation reduces to a dot product of the weight and limb darkening profile vectors multiplied
by the area occluded by the planet.

\section{Implementation}
\label{sec:implementation}
\subsection{Basic implementation}

While Eq.~\eqref{eq:model} can be used to directly evaluate the transit model, some simple optimisations can be used to
significantly decrease the computational cost of the model in real-life use cases. In the most basic typical case a single transit model
evaluation considers a single stellar brightness profile, a single radius ratio, and up to hundreds of thousands of star-planet
separation $c$  values. The computation of the weight array separately
for each $c$ would be rather expensive, and can be avoided by precomputing it into a 2d weight vector interpolation table 
for a set of grazing values $g = c/(1+k)$ for $g = [0..1]$. I parameterise the array with a grazing value 
rather the star-planet distance $c$ because this simplifies the interpolation in the case we want to precompute the 
weight array also as a function of $k$, giving us a 3d weight vector interpolation table, and then interpolate the weight 
vector in the $(k,g)$-space, as detailed later in Sect.~\ref{sec:implementation.interpolation}.

\subsubsection{Interpolation in $g$}

First, the stellar disk is discretised into concentric annuli (bins) during the model initialisation. The model uses a discretisation
strategy where the stellar disk is divided into an inner and outer region divided at $z_\mathrm{cut}$, as shown
in Fig.~\ref{fig:ptmodel}. The bin width is constant in $z$ in the inner region 
and constant in $\mu$ in the outer region.  The use of bins with constant $\Delta z$ 
in the inner region ensures that the central regions of the stellar disk are sampled well, and the use of bins with constant $\Delta \mu$ 
in the outer region ensures that the limb of the star where the surface brightness changes rapidly is also sampled well.
The inner and outer region resolution and the dividing $z$ value can be optimised automatically given a limb darkening model 
and an allowed error threshold. 

The discretisation yields two arrays storing the $z$ values of the bin edges and means, and an array containing the $\mu$ values 
corresponding to the mean $z$ values can be calculated for the limb darkening model evaluation since the limb darkening models are generally 
expressed as functions of $\mu$. In \python, the model is initialised as
\begin{lstlisting}[label=lst:grid_creation]
ze, zm = create_z_grid(zcut, nzin, nzlimb)
mu = sqrt(1 - zm*zm)
\end{lstlisting}
where \texttt{zcut} is the radius that defines the inner and outer areas of the stellar disk, and
\texttt{nzin} and \texttt{nzlimb} the inner and outer region resolutions.

After the initialisation, the model can be evaluated for an array of planet-star centre distances, \texttt{c}, given a radius 
ratio, \texttt{k}, limb darkening model \texttt{ldmodel}, limb darkening model parameters, \texttt{ldpar}, the integrated stellar 
brightness, \texttt{istar}, and the $g$-discretisation resolution, \texttt{ng}, as
\begin{lstlisting}[label=lst:model_evaluation]
dg, w = calculate_weights(k, ze, ng)
ldp = ldmodel(mu, ldpar)
ldw = dot(w, ldp)

for iz in range(nz):
    g = z[iz] / (1+k) 
    i = int(floor(g/dg))
    a = g/dg - i
    ip = (1-a)*ldw[i] + a*ldw[i+1]
    ap = cc_intersection_area(k, z)
    f[iz] = (istar - ip * ap) / istar
\end{lstlisting}
The model evaluation will be rather slow if implemented in pure \python. However, the speed will  be close to \textsc{C} of \textsc{Fortran}
performance if the loop and all the main functions are accelerated using \textsc{numba}.

\subsubsection{Interpolation in $k$ and $g$}
\label{sec:implementation.interpolation}
The previous approach calculates the weight array for each new planet-star radius ratio $k$. This can be avoided by precomputing
a 3D weight array during the model initialisation so that the weight vector can be interpolated as a function of $k$ and $g$. For this, 
we need to give the model the minimum and maximum radius ratio limits and the radius ratio resolution. Now, the weight calculation is
moved to the model initialisation
\begin{lstlisting}[label=lst:grid_creation]
ze, zm = create_z_grid(zcut, nzin, nzlimb)
dk, dg, w = calculate_weights_3d(nk,k0,k1,ze,ng)
mu = sqrt(1 - zm*zm)
\end{lstlisting}
and the only required modification to the model evaluation is one additional interpolation
\begin{lstlisting}[label=lst:model_evaluation]
ldp = ldmodel(mu, ldpar)
nk = (k - k0) / dk
ik = int(floor(nk))
ak = nk - ik
ldw = (1-ak)*dot(w[ik],ldp) + ak*dot(w[ik+1],ldp)

for iz in range(nz):
    ...
\end{lstlisting}

While promising in theory, the performance gain offered by interpolating in $k$ and $g$ is rather insignificant what comes
to the total evaluation speed of the transit model when including also the Keplerian orbit computations, etc. Both approaches
are implemented in \pytransit, interpolation in $g$ only (from here named as \textit{direct} model) is the default setting,
and the model using $(g,k)$ interpolation (from here named as \textit{interpolated} model) can be switched on in the model initialisation easily.

\subsection{Limb darkening}
\label{sec:implementation.limb_darkening}
\subsubsection{Named limb darkening models}
The limb darkening model is chosen in the transit model initialisation, and can be one of the
named limb darkening models built in to \pytransit or a Python callable. The buit-in models are\footnote{See
\citet{Mandel2002}, \citet{Gimenez2006}, \citet{Parviainen2015b}, and \citet{Kreidberg2015} for overviews of the analytical forms. 
The \textit{triangular quadratic} is an alternative parametrisation to the quadratic model by
\citet{Kipping2013b}, and the \textit{power-2-pm} model is an alternative parametrisation to the power-2 
model by \citet{Maxted2018}.}
\begin{itemize}
    \item uniform
    \item linear
    \item quadratic
    \item triangular quadratic
    \item general
    \item nonlinear
    \item logarithmic
    \item exponential
    \item power-2
    \item power-2-pm
\end{itemize}
and are all integrated analytically over the stellar disk to get an exact normalisation factor. 

\subsubsection{Custom limb darkening model}
The \swift model can model limb darkening by any \textsc{Python} callable that returns an array of surface brightness values when 
given an array of $\mu$ values and a parameter vector. If the model is given a single callable rather than an LD model 
name in the initialisation, the normalisation factor is integrated numerically with a minor performance loss. If the
model is given a tuple containing two callables, the second one is assumed to calculate the normalising integral in an 
optimised fashion for a given set of LD parameters.

\subsubsection{\textsc{LDTk} limb darkening model}
Since the limb darkening is not restricted to be an analytical function of any sort, we can, for example, use
numerical limb darkening profiles created by stellar atmosphere models directly. \pytransit takes advantage
of this possibility and offers an \ldtk{}-based \citep{Parviainen2015b} limb darkening model to be used with the \swift transit model.
The \ldtk limb darkening model creates a sample of limb darkening profiles given a set of transmission functions 
defining the passbands and a set of stellar parameters and their uncertainties. Each model evaluation draws a 
random profile from the sample and interpolates the brightness values for the given $\mu$ values from the tabulated 
specific intensity spectra. 

The \ldtk limb darkening model uses PHOENIX-calculated stellar spectra by \citet{Husser2013}, but the approach can be 
extended to use any stellar spectra calculated for a set of $\mu$ values. The approach of using modelled stellar 
spectra directly is similar to setting \ldtk-calculated priors on the limb darkening coefficients but we do not
need to worry about how well the chosen LD model can reproduce the true LD profile, and the final model will
have a smaller number of free parameters since the limb darkening is parameter (coefficient) free. However,
introducing additional uncertainty (to account for possible biases in the stellar atmosphere models) becomes difficult. 

\subsection{Accuracy}
\begin{figure*}
	\centering
	\includegraphics[width=\textwidth]{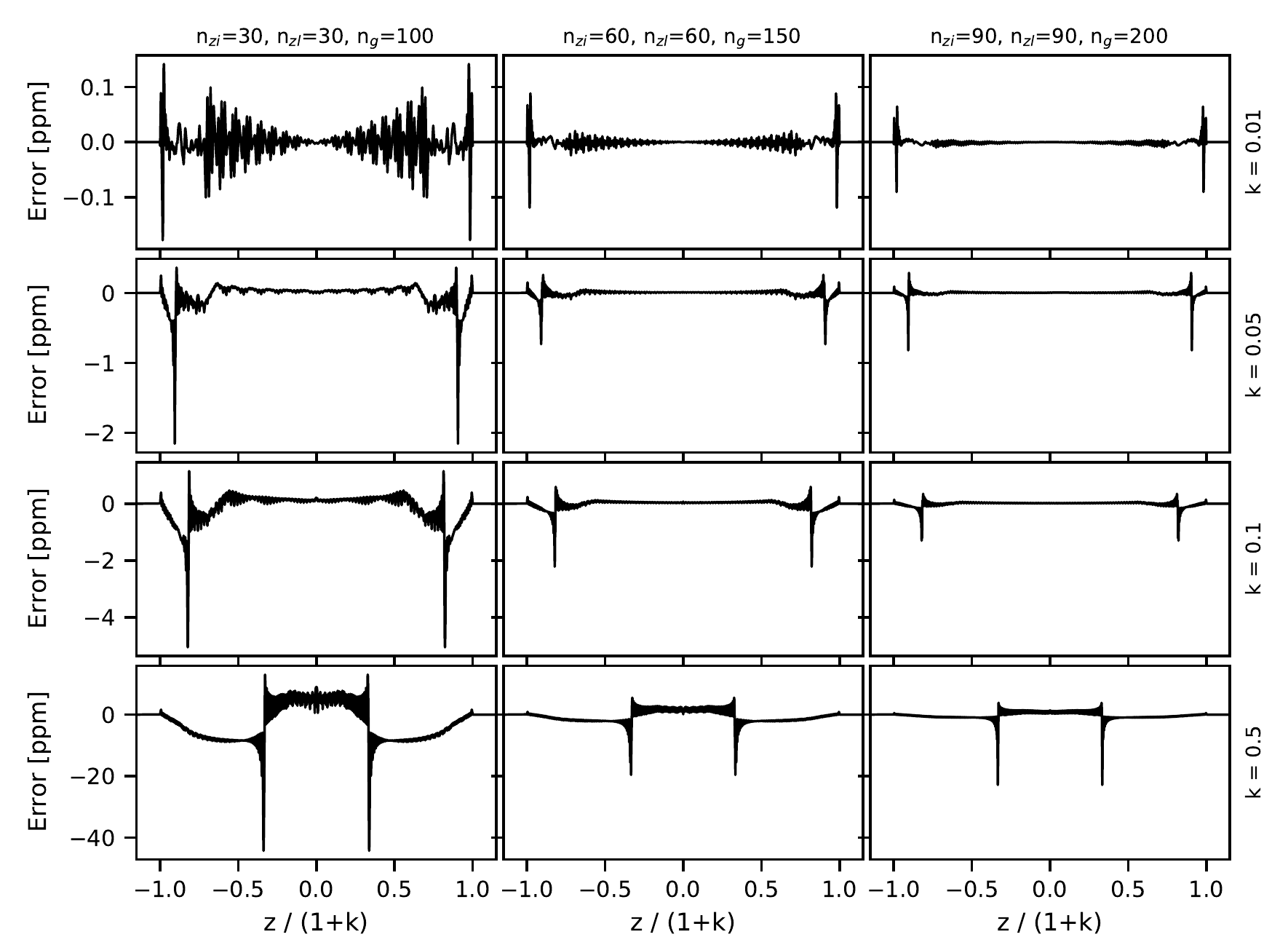}
	\caption{The \swift model error as a function of the grazing value ($z / (1+k)$) in parts per million 
	for quadratic limb darkening for $k$ values of 0.01, 0.05, 0.1, and 0.5, $n_\mathrm{z}$ of 60, 12, and 18,
	and $n_\mathrm{g}$ of 100, 150, and 200. The y-scale is the same for all columns in a single row, but 
	varies from row to row. The error is computed against the analytical quadratic limb darkening model by
	\citet{Mandel2002}.}
	\label{fig:error}
\end{figure*}

The accuracy of the model can be tuned by modifying the $g$, $z_\mathrm{in}$,
and $z_\mathrm{limb}$ resolutions and the $z_\mathrm{cut}$ location. The model shows
no asymptotic biases and a sub-ppm error level can be achieved while still keeping a practical performance
(although this sort of accuracy is unlikely necessary in the near future). \change{Figure~\ref{fig:error} shows the
direct model error as a function of the grazing value for quadratic limb darkening for
four radius ratios and three integration resolution setups. The errors are computed against the analytic
quadratic limb darkening model by \citet{Mandel2002}. Since the lowest-resolution setup already yields
maximum errors in the range of 5~ppm, the resolution can still be significantly reduced for most current
scientific use cases.}

\subsection{Optimisations}

\subsubsection{Selection of the implementation}

Many variables outside the basic computation algorithm can affect the execution speed of the code. The processor, operating 
system, and size of the simulated dataset can all affect whether, for example, parallel or serial computation is the optimal 
approach. Multithreading can increase the execution speed with modern multicore computers significantly, but 
comes with initialisation overheads that can decrease the overall evaluation speed if the problem is not large enough to
justify them. 
In our case, because the transit model is very cheap to compute, the per-call parallelisation overhead can be 
many times larger than the actual transit model computation time if the number of points to evaluate the model at is small 
(smaller than some thousands of in-transit datapoints).

\pytransit implements several versions of the model (parallel, serial, simplified) and chooses the best implementation
after the model has been set up by making a quick trial run with each and selecting the fastest. This is invisible to the
end user, generally takes some fractions of a second, and should ensure that the optimal model is chosen always.

\subsubsection{Small-planet approximation}
If the planet is small enough, the weighted average can be replaced with the limb darkening value evaluated at the centre of
the planet. The small-planet approximation radius ratio threshold can be set at the model initialisation.

\subsubsection{Transmission spectroscopy}
The passband-dependent variations in the radius ratio are small compared to the average radius ratio and have an insignificant
effect on the limb darkening weights. Thus, for transmission spectroscopy, we can use a single limb darkening weight array
calculated for the mean radius ratio rather than calculating the weight array separately for each passband. This approach
reduces the computation cost of the model for transmission spectroscopy.

\subsubsection{GPU computation}
The \swift transit model evaluation is so computationally cheap that using a GPU just for the evaluation of a single transit model
does not bring forth any significant speed benefits, and can actually be slower due to the need to copy data between the main
memory and the GPU. However, GPU evaluation can increase the evaluation speeds significantly (10-20 times) when we increase the
computational burden of a single datapoint evaluation (such as when using supersampling), and when we want to evaluate the model 
for a large set of parameters simultaneously, such as when using \emcee \citep{Foreman-Mackey2012} for Markov Chain Monte Carlo sampling. 

\pytransit implements the \swift model for GPUs using \textsc{OpenCL}, and the GPU version can be used as a drop-in replacement 
for the CPU implementation without any modifications. However, the GPU version is limited to 32 bit floating point precision
and may not be suitable when extreme precision is required.

\section{Model usage}
\label{sec:usage}
\subsection{Basic usage}
The model is prepared for use by initialising an instance of the \texttt{RoadRunnerModel} class and giving the model a set of 
arrays defining the modelled light curves
\begin{lstlisting}[label=lst:model_initialisation]
from pytransit import RoadRunnerModel
tm = RoadRunnerModel('power-2')
tm.set_data(times)
\end{lstlisting}
The \cp{RoadRunnerModel.set\_data} method takes at minimum an array of mid-exposure values to evaluate the model at, 
here given in the \cp{times} array, but it can also be given the model supersampling factor and exposure time of individual
exposures when modelling long-cadence observations \citep[from \textit{Kepler} or \textit{TESS}, for example,][]{Kipping2010a}
\begin{lstlisting}[label=lst:model_initialisation_ss]
tm.set_data(times, nsamples=10, exptimes=0.02)
\end{lstlisting}
The model is ready to be evaluated after the setup
\begin{lstlisting}[label=lst:model_evaluation]
flux = tm.evaluate(k, ldc, t0, p, a, i, e, w)
\end{lstlisting}
where \cp{k} is a float or a 1D or 2D array containing radius ratios, \cp{ldc} is a 2D array containing limb
darkening parameters (coefficients), and \cp{t0},  \cp{p}, \cp{a},  \cp{i}, \cp{e},  \cp{w} are either floats or 1D 
arrays containing the orbital parameters (zero epoch, period, normalised semi-major axis, inclination, eccentricity,
and argument of periastron, respectively).

If the radius ratio and orbital parameters are floats, the model is evaluated for this set of scalar parameters, and the
returned flux array will be one-dimensional with a shape $(n_\mathrm{pt})$, where $n_\mathrm{pt}$ is the number of
mid-exposure times to evaluate the model at. However, if the radius ratios and orbital parameters are given as 
one-dimensional arrays of size $n_\mathrm{pv}$, the model will be evaluated for all the given parameters in parallel,
and the returned flux array will be two-dimensional with a shape  $(n_\mathrm{pv}, n_\mathrm{pt})$. The parallelised
evaluation can lead to very significant performance boost when using a global optimisation or MCMC method
that benefits from parallelised code.

\subsection{Heterogeneous light curves}

The mid-exposure times are the minimal amount of information required by the model, but more can be given when
modelling heterogeneous light curves. Now, the modelled dataset consists of several light curves with (possibly) 
different passbands, exposure times, and required supersampling rates, and the radius ratio and limb darkening
coefficients can be treated as passband-dependent variables. The full form of \cp{RoadRunnerModel.set\_data} is
\begin{lstlisting}[label=lst:model_initialisation_ss]
tm.set_data(times,lcids,pbids,nsamples,exptimes)
\end{lstlisting}
where \cp{lcids} is an integer array of size $n_\mathrm{pt}$ containing the per-exposure \textit{light curve indices} 
that map each exposure into a a single light curve, \cp{pbids} is an integer array of size  $n_\mathrm{lc}$ containing
the per-light-curve \textit{passband indices} that map each light curve into a single
passband, where $n_\mathrm{lc}$ is the number of separate light curves, and \cp{nsamples} and \cp{exptimes} can be
also given as arrays setting the oversampling rate and exposure time separately for each light curve.

As a dummy example, a heterogeneous time series consisting of three light curves observed in two different passbands
would be set up as
\begin{lstlisting}
times = [0, 1, 2, 3, 4, 5]
lcids = [0, 0, 1, 1, 1, 2]
pbids = [0, 1, 0]
tm.set_data(times, lcids=lcids, pbids=pbids)
\end{lstlisting}
After the setup, the model can be evaluated either assuming a achromatic (passband-independent) radius ratio
\begin{lstlisting}
tm.evaluate(k=0.10, 
            ldc=[[0.2,0.1, 0.5,0.1]], 
            t0=0.0, p=1.0, a=3.0, i=0.5*pi)
\end{lstlisting}
chromatic (passband-dependent) radius ratio
\begin{lstlisting}
tm.evaluate(k=[0.10, 0.11], 
            ldc=[[0.2,0.1, 0.5,0.1]], 
            t0=0.0, p=1.0, a=3.0, i=0.5*pi)
\end{lstlisting}
or either but for a number of parameter sets simultaneously
\begin{lstlisting}[label=lst:model_evaluation]
tm.evaluate(k=[[0.10, 0.12], [0.11, 0.13]],
            ldc=[[0.2, 0.1, 0.5, 0.1],
                 [0.4, 0.2, 0.6, 0.2]],
            t0=[0.0, 0.01], 
            p=[1.1, 1.2], 
            a=[3.0, 2.9], 
            i=[0.5*pi, 0.49*pi])
\end{lstlisting}
where in the last case \cp{k} is a 2D array of shape $(n_\mathrm{pv}, n_\mathrm{pb})$, the limb darkening parameters
(\cp{ldc}) are given as a 2D array of shape $(n_\mathrm{pv}, n_\mathrm{pb}\times n_\mathrm{ldc})$,
$n_\mathrm{pb}$ is the number of passbands, and $n_\mathrm{ldc}$ is the number of limb darkening model parameters .
I have omitted the eccentricity and argument of periastron above because they are optional and default to zero
(circular orbit) if not given.

\subsection{\textsc{LDTk} model}
The \ldtk-based limb darkening model presented in Sect.~\ref{sec:implementation.limb_darkening} can be set up with
minimal effort by initialising it with the passbands, stellar parameters and their uncertainties, number of samples to use,
and whether to initially "freeze" the model (assuming \ldtk is installed, of course).
\begin{lstlisting}[label=lst:model_initialisation]
from pytransit import RoadRunnerModel, LDTkLD
from ldtk import sdss_r, sdss_i

ld = LDTkLD((sdss_r, sdss_i),
            teff=(3425, 50), 
            logg=(4.6, 0.1), 
            z=(0.0, 0.1),
            samples=1000,
            frozen=True,
            lowres=True)
            
tm = RoadRunnerModel(ld)
tm.set_data(...)

<MODEL OPTIMISATION>

ld.thaw()

<MCMC SAMPLING>
\end{lstlisting}
A frozen LD model returns values from the mean limb darkening profile rather than a profile drawn randomly from the sample.
This is important during the model optimisation since a randomly drawn LD profile would confuse the optimiser. After the optimisation
and before the MCMC sampling, however, the model should be released (thawed) to marginalise over the limb darkening profiles.

The final option (\cp{lowres=True}) tells \ldtk to use light-weight low-resolution spectra binned from the original \citet{Husser2013}
models to 5~nm spectral resolution. This should suit most exoplanet transit modelling needs but reduces the 
sizes of the downloaded spectra to 2\% from the original.

\section{Performance}
\label{sec:performance}

\begin{figure*}
	\centering
	\includegraphics[width=\textwidth]{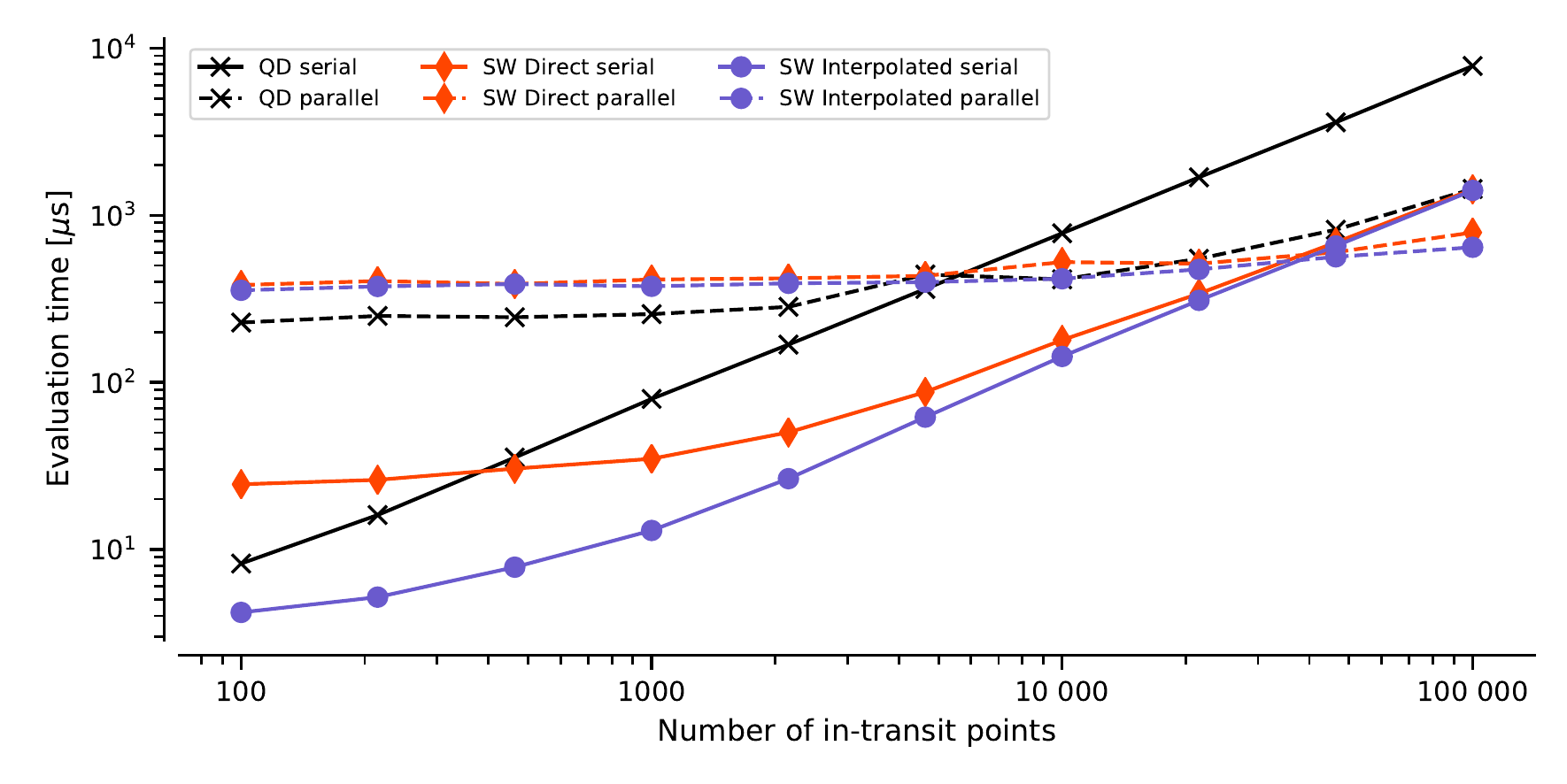}
	\caption{Evaluation time for different versions of the \swift transit model and the quadratic model
	by \citet{Mandel2002} as a function of model in-transit points to calculate. The \swift model is 
	calculated with an accuracy that gives maximum absolute error of 9~ppm and mean absolute error of 0.3~ppm.}
	\label{fig:performance}
\end{figure*}

I show the performance of the different \swift transit model implementations and an optimised quadratic model 
in Fig.~\ref{fig:performance}. The direct serial \swift model is faster than the quadratic model when the
number of in-transit points is larger than $\sim$400, and the interpolated \swift model is faster always. 
The interpolated model offers an advantage over the direct model when the number of in-transit points is
smaller than $10^5$. After this, the model evaluation time dominates the total evaluation time, and the
additional per-call time to calculate the limb darkening tables becomes irrelevant. However, the interpolated
model should never be slower than the direct model, and it's only disadvantages are a small reduction in 
accuracy and the necessity to know the allowed minimum and maximum radius ratios.

Threading initialisation adds a significant overhead to the parallelised implementations and the parallelised
versions are faster than the serial ones only when we reach tens of thousands of points. However, this
applies only to the evaluation of the transit model itself, what is relatively cheap for both the \swift
and quadratic model. Parallelising the whole light curve calculation including the calculation of the normalised
planet to star distance for eccentric orbits and supersampling increases the computational burden for a single
modelled photometry observation, and benefits from parallelisation (either multithreading or GPU computation)
significantly earlier.\!\footnote{I am not showing the performance tests considering parallelising the whole 
light curve computation here because a) it is the same for all transit models, and b) it would deviate from the
main focus of the paper.}

\section{Conclusions and Discussion}
\label{sec:conclusions}

I have presented \swift, a fast exoplanet transit model that can be used with any radially symmetric limb
darkening model. The CPU and GPU implementations of the model available in \pytransit~2.1 have been tested 
thoroughly and offer an unified interface that is simple but still flexible enough to model complex heterogeneous 
time series observed in different passbands, and supersampling requirements. The model offers similar or higher
performance to the analytical quadratic transit model by \citet{Mandel2002}, but allows complete flexibility 
to what comes to the stellar limb darkening. The model can use numerical limb darkening models just as well as 
analytical ones, what allows it to be paired, for example, with \ldtk to use limb darkening profiles created
by stellar atmosphere models directly.

The model already performs well, but future may show further possibilities for optimisation and improvement.
Especially, while the \textsc{OpenCL} implementation of the \swift model is fully functional and can accelerate
the computation speed 10-20 fold, its performance can likely be improved further. Also, the \textsc{OpenCL} 
implementation is limited to 32~bit floating point precision, which means that it cannot be directly used with
observations requiring ppm precision. Some mixed CPU and GPU approaches could however allow to overcome
the precision issue while still offering an improvement in the model performance (for example, calculating the
normalised distances and the $\hat{I}_\mathrm{p} A_\mathrm{p}$ term in the GPU and averaging over the latter
term over the subsamples), but I leave the studies whether they are practical or not to the future.

The \pytransit package containing the implementation of the \swift model has been under continuous development 
since 2010, and contains CPU implementations of six transit models and GPU implementations of four, all with
an unified application interface (API). Version 2 moved to use \textsc{numba}-accelerated model implementations 
over the \textsc{Fortran} implementations of \pytransit~1, which allowed for easy installation in every
\textsc{Python}-enabled computation environment and platform-independent parallelisation. The near future of 
the package will contain further stability testing to ensure that all the implemented transit models work robustly,
and the slightly further future will see performance and usability optimisations, especially focusing on
improving the performance of the GPU implementations.

\section*{Acknowledgements}
I thank the anonymous referee for their helpful and constructive comments.
I acknowledge financial support from the Agencia Estatal de Investigación del Ministerio de Ciencia, Innovación y
Universidades (MICIU) and Unión Europea Fondos FEDER (EU FEDER) funds through the project PGC2018-098153-B-C31.

\section*{Data availability}
There are no new data associated with this article.

\bibliographystyle{mnras}
\bibliography{ptmodel} 


\bsp
\label{lastpage}
\end{document}